\documentclass[aps,prd,twocolumn,groupedaddress,nofootinbib,showkeys]%
{revtex4-1}

\usepackage[T1]{fontenc}
\usepackage{enumerate}
\usepackage{bm}
\usepackage{amsmath}
\usepackage{amssymb}
\usepackage{graphicx}

\newcommand{\scrif}{{\mathcal{I}^{+}}}

\begin{document}

\title{Hawking radiation, quantum fields, and tunneling}

\author{Adam D. Helfer}

\email[]{helfera@missouri.edu}
\affiliation{Department of Mathematics and Department of Physics \& Astronomy,
University of Missouri,
Columbia, MO 65211, U.S.A.}

\date{\today}

\begin{abstract}
I consider the problem of reconciling ``tunneling'' approaches to black-hole radiation with the treatment by quantum field theory in curved space--time.  It is not possible to do this completely, but using what appears to be the most direct and natural correspondence, the simplest case of such an approach does not describe a tunneling process but rather a WKB approximation to a two-point function.  What it computes is
a sort of rescaled Unruh temperature associated with world-lines tracing Killing trajectories near the horizon.  This temperature is numerically equal to the Hawking temperature, but at this point no argument is known which identifies it with the Hawking effect; indeed, the same temperature can exist when no curvature is present.
\end{abstract}


\maketitle

\section{Introduction}
\label{sec:intro}

Hawking's prediction that black holes radiate \cite{Hawking1974,Hawking1975} remains one of the intriguing facets of relativity.  It was initially received skeptically, but has come to be embraced by virtually all workers in the field.  However, while there are innumerable enthusiasts for the prediction, there is decidedly less agreement about what its foundations are.  

Hawking's original argument, which was based on a careful and insightful analysis of the propagation of quantum fields on a gravitationally collapsing space--time, revealed also that trans-Planckian modes would enter essentially.  
He pointed out too he had neglected possible quantum-gravitational effects, and that these might alter his predictions; and indeed it was later shown
on dimensional grounds that quantum-gravitational effects could 
very plausibly completely change the picture (though they certainly need not).  

These two problems are simply there.  They are {\em consequences} of conventional quantum field theory in curved space--time, and so no derivation of Hawking radiation {\em within}, or even {\em strictly compatible with}, that framework can avoid them.  They would have to be overcome by essentially new physical hypotheses.\footnote{There is an exception to this.  If one introduces a textbook ultraviolet cut-off, there is no trans-Planckian problem, but also no Hawking radiation.  There is no absolute reason to reject this possibility, although it is hard to let go of the very beautiful link Hawking's work suggests between black holes and thermodynamics.}  (See ref. \cite{ADH2003} for a review of these matters.)

But it is not easy to see how quantum field theory in curved space--time should be modified.  
Quantum field theory itself was largely developed in order to reconcile quantum theory with special relativity, and it has been extraordinarily successful where it has been tested.  The transition to quantum field theory in curved space--time is based largely on the hypothesis that the local structure of the theory should accord with the Minkowskian one.  So it is hard to see how quantum field theory in curved space--time could be changed without upsetting either special-relativistic quantum theory or local Lorentz invariance.

The seriousness of the problems underpinning Hawking radiation makes it desirable to search for alternative derivations, but the hermetic character of the arguments for quantum field theory in curved space--time makes this very difficult.  
I will be concerned here with one family of ideas to treat black-hole radiation, the ``tunneling'' proposals.  

\subsection{``Tunneling''}

The first ``tunneling'' proposal was
from Parikh and Wilczek \cite{PW2000}; a somewhat different (``Hamilton-Jacobi'') form was given by Angheben et al.  \cite{Angheben_2005} and is most commonly used now.  See ref. \cite{Vanzo_2011} for a review.
The proposals are supposed to give a treatment of black holes somewhat parallel to the Schwinger effect \cite{Schwinger1951} (which predicts the creation of charged pairs by a strong electric field\footnote{This prediction was actually made earlier, by Euler and Heisenberg \cite{Heisenberg_1935}, before the modern development of quantum electrodynamics.}).

The physical interpretation of these proposals, however, remains obscure.  This is because, while much suggestive 
language (tunneling, virtual particles, ...) has been used, it has so far not been possible to establish detailed 
models or justifications for all the steps.  
It is not in fact clear at this point whether these ideas are supposed to be consequences of quantum field theory in curved space--time, or alternatives to it --- and, if they are alternatives, what their foundations are.

A great deal of this is because of
unresolved usages of the term ``particle.''  
Hawking {\em based} his analysis on a careful discussion of what one should reasonably call physical particles.  
However, it is not clear how much of this is accepted in the tunneling papers, and 
straightforward readings of many of them conflict strongly with Hawking's.

The papers invoking tunneling generally present the process as the creation of ``particles''  infinitesimally outside of the event horizon, which then simply suffer red-shifts (and have amplitudes somewhat reduced by greybody factors) as they propagate to future null infinity $\scrif$.  If this were really the case, then the particles created near a given point on the horizon would be blue-shifted relative to the Hawking temperature, by amounts exponentially increasing with retarded time (in any fixed frame near the horizon).   
Colossal energy densities would be present, diverging at the horizon. 

This is not at all what Hawking's analysis showed, nor indeed what will happen in any analysis consistent with quantum field theory in curved space--time.  In those pictures, physical particle production is a highly non-local matter, and one can only unambiguously say that Hawking quanta are present at distances greater than a few gravitational radii from the hole.
Closer to the horizon, there is no unambiguous physical definition of particles of the relevant wave-vectors, and local measurements of the physical state would reveal no remarkable structure:  the Hawking quanta are present only in the sense that they will be revealed by a Bogoliubov transformation from local field data there to the neighborhood of $\scrif$.

It is acknowledged in some of the tunneling literature that the term ``particle'' is used in an extended sense and that it is coordinate- (or frame-) dependent \cite{Vanzo_2011}.  Presumably, it is hoped that eventually one will be able to develop a theory of just what these ``particles,'' and their physical properties, are, so that one will be able to give a coherent account of how and where they are created, how they propagate, and what their stress--energy content is.  But these things have yet to be done.

So we await clarifications of in just what sense particles are supposed to be produced at the horizon, and just how they are affected as they move outwards to $\scrif$, in the tunneling picture.

Generally, the papers compute a quantity they refer to as the imaginary part of the action $\Im S$, and, usually without discussion, identify this with particle production. 
It is true that one can relate the particle production Schwinger found to what might be called the imaginary part of the action for an auxiliary problem, but this identification strongly takes advantage of the Minkowskian background structure.  One needs to justify a parallel application to the gravitational case.

It should also be emphasized that,
while the language of tunneling is often used in connection with the Schwinger effect, it is not really a correct description of that process. 
This description invokes the ``Dirac sea'' in a way which is considered in modern field theory not just to be incorrect but to be a source of much 
confusion.\footnote{Schwinger himself wrote:
\begin{itemize}
\item[]
The picture of an infinite sea of negative energy electrons is now best regarded as an historical curiosity, and forgotten.  Unfortunately, this episode, and discussions of vacuum fluctuations, seem to have left people with the impression that the vacuum, the physical state of nothingness (under controlled physical circumstances), is actually the scene of wild activity. \cite{Schwinger1973}
\end{itemize}
The first sentence is quoted in Weinberg's book on quantum field theory \cite{Weinberg1995v1}.  Zee says (of particles with negative energies, and traveling backwards in time) this ``metaphorical language, when used by brilliant minds... was evocative and inspirational, but unfortunately confused generations of physics students and physicists'' \cite{Zee2003}.
}
So even if one could present the Hawking effect as a parallel to a tunneling calculation for the Schwinger effect, one would have a lot of explaining left to do.
In fact, Hawking himself considered the possibility of a tunneling description, and cautioned it was ``heuristic only and should not to be taken too literally'' \cite{Hawking1975}. 

We are left with an unsettled situation.  
The ``tunneling'' approaches depend on a formula taken 
over from special-relativistic quantum electrodynamics,
but there are questions about the justification for this in the gravitational case, and it is not clear if the results are compatible with quantum field theory in curved space--time.  

\subsection{This paper}

The aim of this paper is to give a natural interpretation of the simplest case of ``tunneling'' calculations  in terms of quantum field theory in curved space--time.
In view of the discussion above, it will not be possible to reconcile all aspects of the tunneling approaches with this theory.  Nevertheless, the present paper at least fills in some elements of one interpretational framework, and provides a point of comparison for others who might wish to offer alternatives.

As the goal here is to treat the simplest case, I will focus on a Hamilton-Jacobi approach.  But the points to be considered here would also appear in a discussion of the Parikh-Wilczek one (although there would be other, complicating, issues).

It will be shown that the 
computations can be viewed as not having to do with tunneling at all, but as giving certain asymptotics for the quantum-field-theoretic two-point functions near the event horizons.  
The temperature computed does indeed correspond to one known from quantum field theory in curved space--time, 
one introduced by Jacobson \cite{Jacobson1996} and closely related to an argument of Unruh's
\cite{Unruh1976}.  And this temperature is numerically equal to the Hawking temperature.  However, it represents a different phenomenon:  it is a kind of acceleration temperature, like that of the Unruh effect.  More precisely, it represents a sort of scaled limit of the acceleration temperature for detectors held at fixed radii, approaching the horizon.

Jacobson suggested applying the principle of equivalence to identify this with the Hawking temperature; however, this is not really feasible, at least in a direct sense.
The difficulty is that
the principle of equivalence only applies in sufficiently small neighborhoods of an event, but the wavelengths of the field modes involved in accelerated-observer effects are necessarily large enough to extend beyond that scale \cite{ADH2003}.
If there is a physically correct argument allowing one to infer the Hawking effect from this scaled acceleration temperature, it has not yet been found.

Another concern is that the mathematical procedure giving this temperature would also (we will see) apply in cases where there is no space--time curvature, for instance in a Rindler wedge.  This strongly suggests that it is simply an acceleration temperature, and not a black-hole radiation one.  

It is worth pointing out that this view is consistent with some of the discussions which have appeared in the ``tunneling'' literature.  As described in ref. \cite{Vanzo_2011}, one can 
try to tackle the question of what ``particles'' are in the ``tunneling'' approaches by 
imagining detectors following some of the coordinate world-lines. 
Here I find that the most straightforward interpretation of the calculations corresponds to a limit of more and more accelerated detectors closer and closer to the horizon.  It is the identification of this with black-hole radiation which is unestablished, and indeed questionable.

So --- with the interpretation adopted here --- the ``tunneling'' approaches compute a sort of scaled limit of acceleration temperatures which is known to be numerically the same as the Hawking temperature, but which we have (so far) no direct way of relating to the Hawking process.

\subsection{Outline}

The next section outlines the ideas leading up to Hawking's prediction of black-hole radiation.  It is not logically necessary for the material which follows (and so can be omitted), but it does help place it in context and may help clarify some of the issues.  The main work of the paper is in Section III, which outlines an interpretation of a simple ``tunneling'' calculation.  The final section is given to discussion.

{\em Conventions.}  The metric signature is $+{}-{}-{}-$; space--time conventions accord with Penrose and Rindler \cite{PR1986}.  Natural units are used, with Newton's constant, the speed of light, Planck's reduced constant and Boltzmann's constant all unity.

\section{Hawking's analysis and its background}

Although the idea that tunneling might be involved in black-hole radiation goes back to one of Hawking's original papers \cite{Hawking1975}, it has remained difficult to make precise arguments supporting it.  In these circumstances, it seems appropriate to give a brief review of the main ways in which concepts suggesting tunneling appeared in Hawking's paper and the work leading up to it.  That is done in this section.  It is not necessary for the rest of the paper, and can be skipped.

In 1971, two papers appeared which showed how energy might be extracted from spinning black holes:  that of Penrose and Floyd \cite{PenroseFloyd} (applying to particles) and that of Zel'dovich \cite{Zeldovich1971} (for fields). 

The Penrose-Floyd argument depended explicitly on the ergosphere, the regime (outside the horizon) in which the Killing vector associated with stationarity became space-like.  Because Killing energy of a freely falling particle is conserved, such a particle might fall into the ergosphere, then split, locally conserving energy-momentum but one of the product particles having {\em negative} Killing energy.  That particle would remain within the ergosphere, but the other could escape with a larger energy than it had initially.  

In this case, for the particle with negative Killing energy, the part of space--time outside of the ergosphere is (classically) forbidden.  This does seem to set up the possibility of a tunneling effect.  However, note that it is associated with the ergosphere and {\em not} the event horizon.

Zel'dovich had argued that a spinning object in Minkowski space could, at the classical level, scatter waves (of some field interacting with the material) in a {\em superradiant} fashion, with the object's rotational energy converted into extra wave energy.  He pointed out that at a quantum level this would amount to stimulated radiation, and that that in turn would imply {\em spontaneous} quantum emission from the object.  He had then asserted that the same would apply to black holes.  When, a little later, he investigated the Kerr case in more detail, his analysis strongly involved the ergosphere \cite{Zeldovich1972}.

Hawking (whose first papers on this appeared in 1974 and 1975) was largely motivated by Zel'dovich's work, although apparently he had reservations about Zel'dovich's quantum treatment \cite{Thorne2004}.  But also Hawking considered an issue which had been neglected by others:  the fact that a hole formed by gravitational collapse would not have been eternal.  This turns out to be critical for the Hawking effect, in contrast to those discussed by Penrose, Floyd and Zel'dovich.  Correspondingly, the Hawking work depends on looking at the propagation of field modes in the neighborhood of the event horizon, whereas the others depend only on penetrating the ergosphere to some degree.

At this point, we see that ideas related to tunneling would have been naturally present in the period leading to Hawking's black-hole radiation work, but also that these ideas are really related to the issue of rotation and not to Hawking radiation.

Hawking's analysis turned on two points:  arguments about what it meant physically to identify particles, and analysis of the propagation of quantum field modes.  The arguments for the physical identification of particles are essentially that if in some space--time volume we can set up a coordinate system in which the curvature scale is $\lesssim L^{-1}$, then we can use the standard forms of the local field operators to sensibly define particles of wavenumbers $\gtrsim L^{-1}$ within the region.  This gives unproblematic definitions of particle content far from the collapsing region.

It was the analysis of the field modes which led to the surprising results.  Ultimately, the definitions of particle number depend on splitting the field modes into positive- and negative-frequency parts, and what Hawking found was that the non-linear red-shift distortion in field modes propagating from the distant past to distant future gave rise to a change in the split for the future regime relative to the past (a Bogoliubov transformation with $\beta\not=0$).  It is this which means that the past vacuum contains future particles.  This distortion is present even in the non-rotating case, and depends strongly on the assumption that the black hole formed from collapse, although it does not depend (after transients) on the details of the collapse.
It is a causal process, and does not depend on penetrating the future event horizon.

Hawking's first, tentative, prediction of black-hole radiation had focussed on the asymptotic regimes, and had shown one would expect a steady emission of thermal radiation.  This could be understood as a quasi-classical energy flux at $\scrif$, but 
if the quasi-classical character was valid everywhere it
would amount to a divergent energy density in the neighborhood of the event horizon.  An essential part of the second paper was to discuss the resolution of this difficulty.

It is there that Hawking presented the argument for the nonlocality of particle creation explicitly and strongly.  This also meant that the field could not be viewed as giving in any direct way a well-defined distribution of particles near the event horizon.  It is in this context that Hawking discussed the possibility of a tunneling interpretation in some detail but ultimately characterized it as ``heuristic only.''

There are a number of issues making a tunneling view difficult to implement it the Hawking-radiation case:

(a) One is looking for tunneling not just across an ergosphere, but across an event horizon, that is, tunneling which overcomes not just an energy barrier, but a causality barrier.  This would strike profoundly at the foundations of relativity.

(b) The Schwinger effect is due to a strong uniform electric field, whereas there need be no locally strong field anywhere in the Hawking picture.

(c) To even begin to formulate a tunneling picture, one would need to discuss its particle-content.  Specifically, one would need to identify a source region, in which the particles are created; outside of this particle-number would be conserved.  But at least the most straightforward approach to this would seem to lead one back to Hawking's analysis and its conclusion that particle production is significantly non-local and does not really occur in any well-defined sense at the horizon.

It is worth adding that
even with a well-defined notion of particles, it
is in general not straightforward to discuss tunneling for relativistic quanta.  The reason is that the concept of tunneling turns on the existence of well-defined states of localization,
so one can meaningfully speak of the amplitude of a particle to go from one place to another.  However, relativistic quanta cannot really be localized below their Compton wavelengths.  

Finally, a comment on issues of localization and the Unruh effect is in order.  Unruh famously showed that a uniformly accelerating detector in Minkowski space would respond as if it were in a thermal bath.  Unruh's model of the detector was essentially confined to a small neighborhood of a central world-line.  Such a detector is best not referred to as a particle detector, since it does not sample a spatial region large compared to the Compton wavelengths of interest.  The detector certainly does respond to the field, however, and one may fairly say it detects field quanta.

\section{Interpreting a ``tunneling'' calculation}

The goal of this paper is to interpret ``tunneling'' computations in quantum-field-theoretic terms.  This leads to a different logical perspective from papers which seek to explore the consequences of such computations.

If one {\em assumes} a tunneling model is valid, then one computes a barrier-penetration amplitude; in the literature, estimates of this turn up a factor controlled by the Hawking temperature, and this temperature is the result of the computation.

Here, however, we wish first to understand what the quantum-field-theoretic basis was for the original tunneling idea --- that is, how the concept of tunneling arose in the Schwinger effect.  Then we ask what would be the parallel in the black-hole case, and then how the actual black-hole ``tunneling'' calculations relate to quantum field theory.  Since the link between the quantum field theory and space--time geometry {\em is} well understood, the main issue will be to 
understand just how the ``tunneling'' calculations select elements of the geometry.

\subsection{The Schwinger effect}

Schwinger considered the effect of a classical electromagnetic field on the electron--positron quantum field.  
He computed the expected current $\langle j_a\rangle$, which one finds from a limit of a two-point function (with a Dirac gamma inserted).  

The core of Schwinger's calculation is finding this two-point function, and this is done by noting that it satisfies a Dirac equation in each variable.  In Schwinger's paper, this appears as an auxiliary to the more fundamental quantum-field-theoretic issues.  However, in the ``tunneling'' approaches, this is described in first-quantized, Dirac-sea, language.  It is said there that negative-energy virtual particles can ``tunnel'' across the negative-to-positive energy mass gap and become real.  A better description is that the electric field contributes enough to the charged field modes' frequencies that a negative-to-positive frequency mixing occurs and what had been a vacuum state, before the external field was applied, becomes a many-particle state.  In other words, we should really use a second-quantized treatment and think in terms of Bogoliubov transformations.

In any event, the key point here is that the ``tunneling'' description is built on calcuating or estimating the two-point function, but describes this estimation in terms of a (not really correct) first-quantized particle dynamics.  
We will try to interpret the ``tunneling'' approach to black-hole radiation into a parallel framework.

\subsection{The black-hole case}

Let $\phi$ be the quantum field, which we take to satisfy the wave equation.  
The quantum state will be $|\Psi\rangle$ and we consider the two-point function
\begin{equation}\label{tpf}
  G(p,q) =\langle\Psi |\phi (p)\phi (q)|\Psi\rangle\, .
\end{equation}
(There is no need to time-order this for our purposes; the time-ordered function could always be recovered from this.)
As the point of the present paper is to offer an interpretation which will at least work in the least problematic situations, we will assume
the space--time is spherically symmetric and represents a spatially bounded gravitationally collapsing body of mass $M$, and shortly we will accept that the dominant contribution to black-hole radiation comes from s-waves.

An obviously correct thing to do would be to compute a weighted average of 
$G(p,q)$, masking $p$ and $q$ by wave-packets approximating Fourier modes near future null infinity, to compute the expected number-density of modes of a given wavenumber there.  This would reproduce results of Hawking's computation, in effect
working out the Bogoliubov coefficients and using explicitly the change in wave-numbers from the distant past (or at least the neighborhood of the horizon) to the neighborhood of future null infinity.  But this is {\em not} what is done in the ``tunneling'' approaches.

There are two central features to the ``tunneling'' approaches:

\begin{enumerate}

\item
They assume we may focus on the ultraviolet asymptotics of the field {\em in the neighborhood of the horizon}.  This is motivated by the correct view that the field modes giving rise to Hawking quanta have ultra-high wavenumbers near the horizon. 

\item
One should resolve the time-dependence in terms of Killing frequencies. 
(It is this step which allows one to write the WKB approximation in a form which will appear as a tunneling calculation.)  
This is obviously mathematically admissible outside the collapsing matter.  However, it does mean disregarding the dynamical, collapse, phase, and the nonlinear distortion of frequencies which, in the Hawking picture, gives rise to the radiation.  That in turn raises the question of just whether the tunneling models are really dealing with black-hole radiation.

\end{enumerate}

Given these two assumptions, one then makes a WKB approximation to solutions of the wave equation.  Let $\xi ^a$ be the standard Killing vector generating the stationarity and normalized at infinity; we may introduce a coordinate $\tau$ such that $\xi ^a\nabla _a \tau=1$.  This coordinate will necessarily be the Schwarzschild time coordinate $t$ plus some function of the Schwarzschild coordinate $r$.  In order to have the coordinate regular across the event horizon, we will choose
\begin{equation}
  \tau = t+r_*
\end{equation}
where
\begin{eqnarray}
  r_*&=& r+2M\log \left[ (r-2M)/(2M)\right]
\end{eqnarray}
is the ``tortoise'' coordinate.  (In fact, $\tau$ is the standard advanced time coordinate, usually denoted $v$.)
Then WKB approximations to s-wave solutions of the wave equation will be superpositions of expressions of the form $\exp i\omega (\tau +\psi (r))$.  The WKB condition applied in the case of the wave equation will force $\tau+\psi$ to be an eikonal.  However, the eikonals are simply functions of $t+r_*$ and functions of $t-r_*$,
so the only non-trivial choice is $\psi =-2r_*$ (up to an additive constant).  The ``action'' $S$ is the change in $\omega\psi$ over a classical path.  The imaginary part of the action will depend only on the whether the end-points of this path are to the left or the right of $r=2M$ and the choice of analytic continuation of the logarithm for $r<2M$.  For a path crossing the horizon once, we have evidently (up to sign)
\begin{equation}
  \Im S =4\pi M\omega \, ,
\end{equation}
and this is the key quantitative result of the ``tunneling'' calculations.  Note that this result would be unchanged under the addition to $\tau$ of any function of $r$ smooth at the horizon.  (In particular, one could work in the Painlev\'e--Gullstand coordinates often used.)

But what have we actually calculated?  What is the interpretation of this in terms of the two-point function?

There is good reason at this point to think that this computation has little to do with gravity.  This is because the main effect comes from moving infinitesimally to first order from one side to the other of a null hypersurface.  If we really were doing this at one point, only the first-order geometry at the point could enter.  In particular, no curvature would be detected, and it would be unclear what the computation had to do with gravity.  

The situation is a bit more complicated, however, because in working with Fourier modes we have really introduced a substantial delocalization --- resolving into these modes requires integrating over the Killing trajectories.  However, since after all the Killing vector generates symmetries it is not clear that this delocalization is enough to bring in very much geometry.  In fact, it is not, as I shall now show.  I will show that the same result for $\Im S$ would be obtained in very general circumstances, including in cases where there is no curvature, for instance accelerated observers in the Rindler wedge.

\subsection{The computation in general}

Let us consider a two-dimensional space--time with a 
Killing field $\xi ^a$ which is null but nonvanishing on a geodesic $N$, timelike future-directed on a set $R$ (for right), and spacelike on a set $L$ (for left).  We may choose null coordinates $U$, $V$, increasing to the future and with $N$ the zero-set of $U$; these coordinates are then unique up to sense-preserving reparameterizations (also preserving $U=0$).  The metric will be $f^2dUdV$ for some positive function $f$.

Whether or not this really describes a black hole, the geodesic $N$ will be a Killing horizon.  We must have
\begin{eqnarray}\label{sgeq}
\nabla _a(\xi ^b\xi _b) = -2\kappa \xi _a\qquad\text{on}\qquad N\, ,
\end{eqnarray}
and the quantity $\kappa$ must be constant on $N$.  In the black-hole case, it would be the surface gravity.  We will assume $\kappa$ is positive.

Let the Killing field be $\xi ^a =\alpha \partial _U+\beta \partial _V$.  Since this must preserve the null geodesics, the coefficient $\alpha$ must depend on $U$ alone, and likewise $\beta$ depends on $V$ alone.  And since $N$ is a Killing horizon, we must have $\alpha (0)=0$.  Since $\xi^a$ is timelike future-pointing, we must have $\alpha>0$, $\beta >0$ on $R$; since $\xi^a\not=0$ on $N$ we must have $\beta >0$ on a neighborhood of $N$.  We may then exploit the reparameterization freedom in $V$ to choose $\beta =1$ near $N$.

A short computation shows that eq. (\ref{sgeq}) implies $\partial _U\alpha =-\kappa\beta =-\kappa$ on $N$.  We may then use the reparameterization freedom in $U$ to choose $\alpha = -\kappa U$.  Then $\xi ^a =-\kappa U\partial _U +\partial _V$, and the Killing condition becomes $f^2 = f^2( V +\kappa ^{-1}\log (-U/U_0))$ for some constant $U_0$, on the set $R$.
(Strictly speaking, it would be better to write $f^2$ as a function of $-(U/U_0)\exp\kappa V$, which would apply across the Killing horizon, but the previous form will fit more directly with subsequent calculations.)

We now choose a coordinate $\tau$ such that $\xi ^a\nabla _a\tau =1$, regular across the horizon.  We may simply take $\tau = V$; as in the previous subsection, it will become evident that the choice, as long as it is regular across the horizon, is irrelevant.  As before, we consider WKB approximations to the wave equation of the form $\exp i\omega (\tau +\psi )$, where $\psi =\psi (V+\kappa^{-1}\log (-U/U_0))$.  Then $\tau +\psi$ must be an eikonal, which is to say a function of $U$ alone or a function of $V$ alone.  The only possibilities are $\psi =0$ and
$\psi =-\kappa ^{-1}\log (-U/U_0) -V$ (up to additive constants).  We see that $\Im S$, identified by looking at the change of $\psi$ across $N$, satisfies (up to sign)
\begin{eqnarray}\label{accS}
\Im S &=& \kappa^{-1} \pi \omega\, .
\end{eqnarray}

This has, of course, the same form as the previous subsection, but all we have assumed here is a Killing horizon.
We see then that this form does not really depend on any gravitational physics at all; it would occur in a Rindler wedge with $\xi ^a=\kappa (x\partial _t+t\partial _x)$ (in standard coordinates) a boost.  

\subsection{Interpretations}

What would be the interpretation of the computation in the Rindler wedge?  We are supposed to be computing some of the ultraviolet asymptotics of the two-point function (holding one point fixed), and we have Fourier-transformed with respect to Killing time.  
Had we Fourier-transformed with respect to proper time $s=(\kappa /a)\tau$, where $a=(x^2-t^2)^{-1/2}$ is the acceleration of the Killing trajectory, we should have expressed the formula (\ref{accS}) as
\begin{eqnarray}\label{accT}
  \Im S = a^{-1}\pi {\tilde\omega}\, ,
\end{eqnarray}
where ${\tilde\omega}$ is the angular frequency with respect to $s$.  We thus see that the formula can be interpreted as recovering the Unruh temperatures $T_{\rm U} = a/(2\pi ) = {\tilde\omega}/(2\Im S)$ of the Killing trajectories, independently of $\kappa$.

On the other hand, we are really interested in this in the limit of small neighborhoods of the Killing horizon, which is to say $a\uparrow\infty$.  In this limit $a$ and the physical Unruh temperature will diverge.  However, we can shift to rescaled parameters $\tau = (a/\kappa )s$, $\omega = {\tilde\omega} \kappa/a$ in terms of which the physical formula (\ref{accT}) has the mathematically stable limit (\ref{accS}).
The role of $\kappa$ in these formulas is simply to provide a reference scale.

Now let us return to the Schwarzschild case.  We may, just as an application of the argument just given, interpret the Schwarzschild computation as simply reflecting the Unruh effect for accelerated observers hovering nearer and nearer the horizon.  In this sense, the mass (and space--time curvature) scale out of the problem.

But there is a further interpretation available in the Schwarzschild case.  We may ask how a distant observer would describe the Unruh radiation in a small box hovering close to the horizon.  The box, and its Unruh temperature, would appear red-shifted.  The acceleration of the box would be
\begin{eqnarray}
  a(r) &=&(1-2M/r)^{-1/2}(M/r^2)\, ,
\end{eqnarray}
with Unruh temperature
\begin{eqnarray}
  T_{\rm U}(r) =  (1-2M/r)^{-1/2}(M/r^2)/(2\pi )\, ;
\end{eqnarray}
the red-shifted temperature was denoted by Jacobson
\begin{eqnarray}
  T_{{\rm U},\infty} (r) =(M/r^2)/(2\pi )\, .
\end{eqnarray}     
As $r\downarrow 2M$, we find this is numerically equal to the Hawking temperature:
\begin{eqnarray}\label{eeeq}
  \lim _{r\downarrow 2M} T_{{\rm U},\infty}(r) =T_{\rm H} =1/(8\pi M)\, .
\end{eqnarray}
We may connect this with the argument of the previous paragraph by noting that the rescaling factor used there was 
\begin{eqnarray}
  a(r)/\kappa = (1-2M/r)^{-1/2} (4M^2/r^2)\, ,
\end{eqnarray}
and this approaches the red-shift factor as $r\downarrow 2M$.

We may summarize these findings as follows.  The computation in the black-hole case amounts to finding the Unruh temperature associated with accelerated detectors near the horizon.  As was noted by Jacobson, if these Unruh boxes are examined by a distant observer, their temperatures will be red-shifted, and in the limit as the boxes approach the horizon the red-shifted temperatures stabilize at the Hawking temperature.

There is, at present, no known correct argument that this equality (\ref{eeeq}) allows one to infer black-hole radiation from Unruh radiation.  The equality is certainly suggestive, but it cannot be considered compellingly remarkable since on dimensional grounds the left-hand side must be a pure number multiple of the right.  For more discussion, see ref. \cite{ADH2003}.

\section{Discussion}

The idea that black-hole radiation might be connected to tunneling goes back to Hawking himself.  Attempts at making this at least semiquantitatively precise go back to Parikh and Wilczek.  But, despite much work on the subject, serious gaps remain in its foundations and interpretation.

The aim of this paper has been to connect at least the simplest sort of tunneling proposal with
conventional quantum field theory.  So it is the s-wave sector of a massless scalar field on a space--time representing the spherically symmetric collapse of a bounded matter distribution to a black hole which is considered, and a Hamilton-Jacobi-type treatment of tunneling.

I have found that an interpretation of the ``tunneling'' calculations in quantum-field-theoretic terms is possible, but it does not quite achieve what proponents of the approach might want.  It does not appeal to tunneling at all; it is better thought of as giving certain estimates of the ultraviolet asymptotics of the two-point function.  

The temperature computed most directly represents a sort of scaled temperature associated with acceleration radiation, for observers hovering closer and closer to the horizon, and was considered earlier by Jacobson.  
It is numerically equal to the Hawking temperature, and Jacobson did suggest trying to appeal to the equivalence principle to identify it with black-hole radiation, but so far no correct argument is known which does this.  The difficulty is that the equivalence principle can only be applied locally, and the wavelengths of the relevant field quanta are too big for the application to be valid.  

Also, the temperature found does not really depend on any space--time curvature being present; it would exist near a Fulling-Rindler horizon in Minkowski space, too.  In this connection, it is worth pointing out that the space--time geometry enters the ``tunneling'' calculation only weakly (through the normalization fo the Killing field).

It remains possible that other interpretations of the ``tunneling'' calculations exist, ones which link them more closely with Hawking radiation.  Whatever interpretations are offered, one would like to know 
just how particles are defined and propagate, and how the proposals are related to quantum field theory.

\bibliography{hrqftprd.reva.bbl}

\end{document}